\documentstyle[aps,prb]{revtex}

\begin{document}
\title{Resonance-like piezoelectric electron-phonon interaction in layered
structures}
\author{B.~A.~Glavin, V.A.~Kochelap, and T.L.~Linnik}
\address{V.~E.~Lashkarev Institute of Semiconductor Physics,\\
Pr.~Nauki 41, Kiev 03028, Ukraine}
\author{A.J.~Kent, N.M.~Stanton, and M.~Henini}
\address{School of Physics and Astronomy, University of Nottingham, Nottingham NG7\\
2RD UK.}
\maketitle

\begin{abstract}
We show that mismatch of the piezoelectric parameters between layers of
multiple-quantum well structures leads to modification of the
electron-phonon interaction. In particular, short-wavelength phonons
propagating perpendicular to the layers with wavevector close to $2\pi n/d$,
where $d$ is the period of the structure, induce a strong smoothly-varying
component of the piezo-potential. As a result, they interact efficiently
with 2D electrons. It is shown, that this property leads to emission of
collimated quasi-monochromatic beams of high-frequency acoustic phonons from
hot electrons in multiple-quantum well structures. We argue that this effect
is responsible for the recently reported monochromatic transverse phonon
emission from optically excited GaAs/AlAs superlattices, and provide additional
experimental evidences of this.
\end{abstract}

\pacs{63.20.Kr, 63.22.+m, 73.63.-b, 73.63.Hs}

\section{Introduction}

It is now well-established that the properties of acoustic phonons in
layered structures can be very different from those in bulk crystals:
mismatch of the elastic properties of the layers results in formation of a
folded phonon spectrum and phonon stop-bands, etc. In particular, this leads
to the reduction of the momentum-conservation restrictions for Raman
scattering of light, which allows probing of the folded acoustic phonons by
light-scattering methods (see Ref.\onlinecite{Cardona} for review). In this
communication we show that a similar effect takes place for piezoelectric
electron-acoustic phonon interactions in layered structures due to mismatch
of the piezo-parameters of the quantum-well and barrier layers. The effect
is especially pronounced for phonons propagating close to the structure axis
and whose wavevector is about $2\pi n/d$, where $d$ is the structure period.
In the following, we will call such phonons {\em resonant}. We show that
the enhanced interaction strongly affects hot-carrier relaxation in multiple-quantum
well (MQW) structures. Namely, {\em spontaneous} emission of collimated
quasi-monochromatic beams of high-frequency phonons by hot electrons becomes
possible. This is in contrast with {\em stimulated} emission of phonons
where their monochromatic character is set due to nonlinear electron-phonon
dynamics in a system with population inversion.\cite{saser} We argue that
recent observation of monochromatic transverse acoustic phonon emission in
GaAs/AlAs multilayered structures\cite{exp-ta} can be attributed to the
considered modificaton of the piezoelectric electron-phonon interaction.
We report also new experimental results that support this interpretation.

\section{Phonon-induced piezoelectric potential in multilayered structures}

To begin, we recall briefly the main features of electron-acoustic phonon
interaction in solitary quantum wells (QW) (see, for example, Ref.%
\onlinecite{Price}). The probability of the electron intra-subband
transitions is proportional to the form-factor, $J$, given by 
\begin{equation}
J\sim \left| \int dz\chi ^{2}(z)\varphi \right| ^{2},  \label{eq:1}
\end{equation}
where $\chi $ is envelope wavefunction of confined electron and $\varphi $
is the phonon-induced perturbation potential. In a uniform medium, phonons
are plane waves and the same applies to $\varphi $. Thus, $\varphi \sim \exp
(iq_{z}z)$, where {\boldmath$q$} is the phonon wavevector and the $z$-axis
is directed perpendicular to the layers. As a result, for $q_{z}\gg 1/d_{QW}$,
where $d_{QW}$ is the QW width, the form-factor is small and 
electron-phonon interaction is suppressed. On the
other hand, interaction with phonons of large in-plane wavevector is
prohibited as well, due to in-plane momentum conservation. These factors
suppress interaction of electrons with high-frequency acoustic phonons. This
holds for both the deformation potential and piezoelectric interaction
mechanisms.

In multi-layered systems, e.g. MQW heterostructures, the situation is
essentially different. The MQW structure is shown schematically in Fig.1.
The QW and barrier width are $d_{QW}$ and $d_{B}$ respectively, and the MQW
period is $d=d_{QW}+d_{B}$. With no tunneling coupling between the adjacent
QWs, the properties of electrons are identical to those in a solitary QW. On
the other hand, the phonons are modified due to mismatch of the {\em elastic}
properties in the QW and the barrier layers. However, for typical MQW
structures this mismatch is small and moderate modifications are noticeable
only in small regions of phonon momentum space corresponding to Bragg
reflections of phonons. This has very little effect on the overall
electron-phonon interaction. However, as we shall see below, mismatch of the 
{\em piezoelectric} parameters, while having negligible effect on the phonon mode
structure, modifies the interaction potential $\varphi$. The latter is
determined by equation 
\begin{equation}
\nabla ^{2}\Phi =\frac{1}{\epsilon \epsilon _{0}}\nabla \cdot %
\mbox{\boldmath $P$}.  \label{eq:2}
\end{equation}
Here $\epsilon $ is the dielectric permittivity and $\epsilon _{0}$ the
absolute dielectric constant. {\boldmath$P$} 
is the polarization induced due to the piezoelectric effect: $%
P_{i}=e_{ijk}u_{jk}$, where $u_{jk}$ is strain tensor and $e_{ijk}$ are
piezo-constants. For the case of strain induced by acoustic phonons of
wavevector $\mbox{\boldmath $q$}=\{q_{x},q_{y},q_{z}\}$, the potential can be
expressed as $\Phi =\phi (z)\exp (i(q_{x}x+q_{y}y))$, and the equation
for $\phi $ is 
\begin{equation}
\frac{d^{2}\phi }{dz^{2}}-q_{\Vert }^{2}\phi =R(z)\exp (iq_{z}z),
\label{eq:3}
\end{equation}
where $q_{\Vert }=\sqrt{q_{x}^{2}+q_{y}^{2}}$ and $R$ is periodic function: $%
R(z)=R(z+d)$. The periodicity of $R$ is due to the piezoelectric 
parameters mismatch in the QW and barrier layers. 
The specific form of $R$ depends on the crystal structure of
the materials which constitute the MQW and also the phonon wavevector and
polarization. It is convenient to expand $R(z)$ as a Fourier series 
\begin{equation}
R(z)=\sum_{n=-\infty }^{\infty }R_{n}\exp (iq_{0}nz),q_{0}=2\pi /d.
\label{eq:4}
\end{equation}
Then, the solution of Eq.~(\ref{eq:3}) is 
\begin{equation}
\phi =-\exp (iq_{z}z)\sum_{n=-\infty }^{\infty }\frac{R_{n}\exp (iq_{0}nz)}{%
(q_{z}+q_{0}n)^{2}+q_{\Vert }^{2}}.  \label{eq:5}
\end{equation}
We see that, for $q_{z}\approx -q_{0}n$, the potential has a component which
is smooth function of $z$. As a result, the form-factor for such $q_{z}$ is
not suppressed, even if $q_{z}a\gg 1$. Furthermore, for small $q_{\Vert }$
the amplitude of the smoothly-varying component is large. Therefore, the
role of mismatch of the piezo-constants is important for resonant phonons
which propagate close to the MQW axis.

Particular characteristics of the piezoelectric interaction are determined
by the coefficients $R_{n}$. It is convenient to present them in the
following form: 
\begin{eqnarray}
R_{0} &=&-q^{2}\frac{e\overline{e}}{\epsilon \epsilon _{0}},  \label{eq:6} \\
R_{n} &=&i\frac{e\overline{\delta e}}{\epsilon \epsilon _{0}}\frac{q}{d}%
\left( 1-\exp (-iq_{0}nd_{B})\right) ,n\neq 0.  \nonumber
\end{eqnarray}
Here $e$ is elementary electric charge and $\overline{e}$ and $\overline{%
\delta e}$ are characteristic piezo-constants, which depend
on the direction and polarisation of phonon. Note that in Eq.~(\ref{eq:6}) and subsequent
equation (\ref{eq:8}) we assume that the amplitude of the displacement for a
phonon is unity). We now consider in detail the example of a GaAs/AlAs MQW
structure with its growth axis parallel to $(001)$ crystal axis. The
structures are composed of cubic-symmetry materials and the piezoelectric
properties are characterized by the only nonvanishing coefficient, $e_{14}$.
Using the isotropic
elastic model we obtain for longitudinal, transverse vertical and transverse
horizontal phonons the following expressions: 
\begin{eqnarray}
\overline{e}^{(LA)} &=&\frac{1}{2}\sin ^{2}\theta \cos \theta \sin 2\phi
_{ph}\left( e_{14}+\frac{d_{B}}{d}\delta e_{14}\right) ,  \label{eq:7} \\
\overline{\delta e}^{(LA)} &=&\delta e_{14}\sin ^{2}\theta \sin 2\phi
_{ph}\left( 1+\frac{3q\cos \theta }{q_{0}n}\right) ,  \nonumber \\
\overline{e}^{(TA,v)} &=&\sin \theta \cos ^{2}\theta \sin 2\phi _{ph}\left(
e_{14}+\frac{d_{B}}{d}\delta e_{14}\right) ,  \nonumber \\
\overline{\delta e}^{(TA,v)} &=&\delta e_{14}\sin \theta \cos \theta \sin
2\phi _{ph}\left( 1+\frac{2q\cos \theta }{q_{0}n}\right) ,  \nonumber \\
\overline{e}^{(TA,h)} &=&2\sin \theta \cos \theta \cos 2\phi _{ph}\left(
e_{14}+\frac{d_{B}}{d}\delta e_{14}\right) ,  \nonumber \\
\overline{\delta e}^{(TA,h)} &=&\delta e_{14}\sin \theta \cos 2\phi
_{ph}\left( 1+\frac{2q\cos \theta }{q_{0}n}\right) .  \nonumber
\end{eqnarray}
Here $e_{14}$ is the piezo-constant in the QW, $\delta e_{14}$ is the
mismatch of the piezo-constants in the barrier and the QW, and $\theta $, $%
\phi _{ph}$ are spherical angles of the phonon wavevector. The polarizations
of transverse phonons are defined such that the displacement for the
horizontally polarized phonon is parallel to the layers of the MQW, and for
the vertically polarized phonon it lies in the plane formed by the
wavevector and the normal to the layers. It is straightforward to obtain
from these equations that close to resonance, $q_{z}=-q_{0}n+\delta q_{z}$, $%
\delta q_{z}\ll |q_{0}n|$, and for small in-plane wavevector, the major
mismatch contribution is due to the transverse phonons, whose form-factors
are 
\begin{eqnarray}
J^{(TA,h)} &\approx &4\frac{e^{2}\delta e_{14}^{2}}{\epsilon ^{2}\epsilon
_{0}^{2}d^{2}}\sin ^{2}\frac{q_{0}nd_{B}}{2}\cos ^{2}2\phi _{ph}\frac{%
q_{\Vert }^{2}}{\left( \delta q_{z}^{2}+q_{\Vert }^{2}\right) ^{2}},
\label{eq:8} \\
J^{(TA,v)} &\approx &4\frac{e^{2}\delta e_{14}^{2}}{\epsilon ^{2}\epsilon
_{0}^{2}d^{2}}\sin ^{2}\frac{q_{0}nd_{B}}{2}\sin ^{2}2\phi _{ph}\frac{%
q_{\Vert }^{2}}{\left( \delta q_{z}^{2}+q_{\Vert }^{2}\right) ^{2}}. 
\nonumber
\end{eqnarray}
As we see, the form-factor dependence on $q_{z}$ has a shape of a peak. As $%
q_{\Vert }$ decreases, the width of the peak decreases and its height
increases. This is illustrated in Fig.~2, where we plot the form-factor for
transverse phonons of vertical polarization and $\phi _{ph}=\pi /4$ as a
function of $q_{z}$ for $q_{\Vert }=10^{7}~m^{-1}$ and $10^{8}~m^{-1}$ (all
form-factors are plotted for $d_{QW}=6~nm$, $d_{B}=2~nm$ and normalized to
its value for $q_{z}=0$; and the form-factor of transverse phonons of
horizontal polarization is similar to that of phonons of horizontal
polarizations). For comparison, the form-factor for a plane-wave potential
is shown in the figure as well. The numbers near the peaks, corresponding to $%
q_{\Vert }=10^{7}~m^{-1}$ indicate the amplitudes of the peak.

It should be mentioned that the described features hold provided the number
of periods in MQW structure, $N$, is big enough. Indeed, as can be shown
from the solution of Poisson's equation for finite-$N$ structure, the resonances
are pronounced if $q_{\Vert }Nd>1$.

These properties of the form-factor allow us to suggest that 2D hot
electrons confined in the QWs of MQW structure can emit quasi-monochromatic
collimated beams of acoustic phonons. To prove this, we performed calculations
of the phonon emission for the experimental setup, which allows to 
register the resonant phonons, see Fig.~3a. In this case the emitted
phonons are registered by the superconducting bolometer deposited on the
back side of the substrate. Typically, the MQW and bolometer lateral
dimensions are much less than the thickness of the substrate. Therefore, the
bolometer probes only phonons propagating close to the MQW axis. 

The results for the spectrum of the phonon emission $Q$ are shown in 
Fig.4. $Q$ is determined as the phonon
power emitted per MQW period per unit area and per unit frequency interval.
The calculations were performed for a GaAs/AlAs MQW where the angle between
the phonon wavevector and MQW axis is restricted by the value $\theta
_{max}=0.1$ (roughly, $2\theta _{max}$ can be estimated as the ratio of MQW
lateral dimensions to the substrate thickness). We used the following
material parameters: density $\rho =5316~kg/m^{3}$, longitudinal and
transverse sound velocities $s_{l}=4730~m/s$, $s_{t}=3345~m/s$,
respectively, $e_{14}=-0.16~C/m^{2}$, $\delta e_{14}=-0.065~C/m^{2}$, and $%
\epsilon =12.9$. These data were taken from Ref.\onlinecite{matdata}. The
temperature of hot electrons was assumed to be $100K$, and the QW and
barrier width are $d_{B}=6~nm$ and $d_{B}=2~nm$. The electron envelope
wavefunctions correspond to confinement in an infinitely deep rectangular
QW. As it was expected, the emission spectrum of the transverse phonons has
sharp peaks for resonant phonon frequencies. For longitudinal phonons, the
contribution of resonant piezoelectric interaction is very much weaker.
Therefore, it cannot be resolved on the broad-spectrum deformation-potential
contribution (in calculations we used deformation potential constant $8~eV$%
). Note, that in experiments the quasi-monochromatic transverse phonon
signal and broad-spectrum longitudinal-phonon signal can be resolved based
on the different travel times necessary to reach the bolometer.

\section{Comparison with exparimental results}

Recently, monochromatic longitudinal and transverse phonon emission was
observed under the femtosecond laser excitation of electrons in GaAs/AlAs
MQW.\cite{exp-la,exp-ta} In these experiments the coherent nature of
longitudinal phonons was detected by high-frequency modulation of the
optical reflection using the pump-probe technique. Complimentary
measurements of the emitted phonon spectral distribution was accomplished
using the phonon filtering method. In additional to conventional structures,
Fig.~2a, ``filtered'' samples were manufactured, where an additional,
mirror, superlattice was grown between the ``generator'' superlattice and
the bolometer. The parameters of the mirror superlattice were selected such
that its stop-bands match the mini--Brillouin zone-center phonon frequencies
of the generator superlattice. Then, the phonon signals for different
excitation photon energies, below and above the fundamental edge of
superlattice, were measured. In the structures without a filter, a
pronounced increase of the phonon signal with increase of the photon energy
was observed, which corresponds to the onset of light absorption in
superlattice. In contrast, for the filtered samples no such increase was
observed for both longitudinal and transverse phonons. This suggests
quasi-monochromatic character of the emitted phonon spectrum. For
longitudinal phonons this result can be attributed to the impulsive
stimulated Raman scattering,\cite{ultrafast-review} which is supported by
the observation of the reflectance modulation at frequency corresponding to
the phonon-miniband center longitudinal phonons. However, no such modulation
is present at the corresponding transverse-phonon frequency. The only
remaining possibility is that monochromatic transverse phonons are emitted
under the relaxation of nonequilibrium carriers following the femtosecond
pulse. However, conventional mechanisms of electron-phonon interaction do
not provide such sharp phonon spectral features. We believe that the
observed monochromatic transverse phonon emission could be attributed to the
resonant piezoelectric-mismatch-induced interaction mechanism proposed in
this paper. Note, however, that the results of the numerical calculations
presented here can not be used for quantitative estimate of the phonon
emission power in these experiments. This is because of complicated
relaxation pattern of electrons and holes in superlattices excited by
ultrafast pulses.

The new results described here concern the angle dependence of the emission.
These were obtained using the phonon imaging technique \cite{Wolfe} in which
the laser spot is scanned across the generator SL, thus changing its
position relative to the detector. We used two SL structures grown by MBE on
0.35 mm-thick semi-insulating GaAs substrates. Sample A contained just a
generator SL and Sample B contained the generator SL and a filter SL. The
generator consisted of 40 periods, each of 22 monolayers (ML) of GaAs and 4
ML of AlAs. For these SL parameters, the first mini-Brillouin zone-centre TA
phonon frequency is 450 GHz. In Sample B, the notch filter was grown below
the generator. It consisted of 40 periods, each of 7 ML GaAs and 7 ML AlAs,
and was separated from the generator SL by a 0.5 micron-thick GaAs spacer
layer. The first mini-Brillouin zone-boundary stop band of the filter is
coincident with the generator frequency and prevents 450 GHz phonons
reaching the detector (bolometer) on the rear face of the substrate. Using
this arrangement, we previously showed that, when the generator SL was
resonantly excited by 100 femtosecond laser pulses, monochromatic TA phonons
of frequency 450 GHz were generated \cite{exp-ta}.

Bolometer signals for on- and off-resonance photoexcitation of the generator
SL in Sample A are shown in Fig. 5.  In this case the laser spot is located
directly opposite the detector.The off-resonance case 
corresponds to the photon energy below the fundamental edge in superlattice but above that 
in bulk GaAs. Here, the bolometer response is due to ballistic and diffusive 
phonons generated via relaxation of the photoexcited carriers in GaAs. 
The on-resonance case corresponds to the photon energy above the fundamental edge in 
the superlattice. Therefore, the increase in signal under resonant
photoexcitation is due to phonon emission from carrier
relaxation in the superlattice, including quasi-monochromatic (450 GHz) phonons. In
the angle dependence measurements, we display the difference between the on-
and off-resonance signals at times close to the ballistic time of flight for
TA phonons (indicated by the dashed lines in Fig. 5). Fig. 6 shows the full
2D image for Sample B (filtered). The pattern is very similar to the TA
phonon focussing pattern in cubic GaAs, convolved with the sizes of the
laser spot (about 40 microns-diameter) and the detector (40 x 40 microns
squared). Linescans, taken horizontally through the centre of the images (as
indicated by the dashed line in Fig. 7), for Sample A and Sample B are shown
in Fig. 7,and clearly show the difference in the angle dependence of the
emission. Is is observed that in the case of the unfiltered sample (A) the
emission is directed into a narrower range of angles. This suggests that the
monochromatic (450 GHz) component of the TA signal is emitted in a direction
close to the SL growth direction as predicted by the theory. Taking account
of the size of the source and detector and also phonon focussing effects, we
can estimate an upper limit for the angle to the SL normal at which the
monochromatic phonons are emitted to be about 10 degrees.

The reported experiments are for femtosecond pulsed optical excitation of
the generator SL. From the above theoretical considerations it follows that
quasi-monochromatic transverse phonons can be emitted under long-pulse
excitation as well. The corresponding measurements are currently in progress
and they will be reported elsewhere.

\section{Conclusions}

In conclusion, we have demonstrated that mismatch of the piezoelectric
constants in the quantum well and barrier layers of multiple quantum well
structures leads to resonance-like enhancement of piezoelectric
electron-phonon interaction for phonons propagating close to the structure
axis and having wavevectror close to $2\pi n/d$. This gives rise to the
emission of quasi-monochromatic phonon beams by hot electrons. We suggest
that this behaviour can account for the recent observation of monochromatic
transverse phonon emission in GaAs/AlAs multiple quantum well structures.
Presented in this paper supplementary measurements of the angle-dependemnce 
of the transverse phonon emission suuport this interpretation. 

The
availability of a source of monochromatic high-frequency transverse phonons
can enhance the appliccability of phonon-spectroscopy methods. As we have
shown, the piezoelectric MQW structures are able to emit such phonons,
potentially, even with the use of relatively simple laser excitation
technique.

We would like to note also that the resonant 
contribution to piezoelectric electron-phonon interaction 
considered here may play essential role
in various vertical transport phenomena in semiconductor superlattice structures.

\begin{acknowledgments}
This work was supported by Royal Society of the UK. BAG nad VAK also acknowledge 
partial support under the STCU grant \#3922.
\end{acknowledgments}


\clearpage
\begin{figure}[tbp]
\caption{ Schematics of MQW structure. Electrons are confined in QW layers
with envelope wavefunction $\protect\chi$. The QW and barrier layers are
characterized by individual values of piezoelectric parameters, $%
\{e^{(QW)}\} $ and $\{e^{(B)}\}$. }
\end{figure}

\begin{figure}[tbp]
\caption{ Form-factors of electron-phonon interaction as a function of $z$%
-component of the phonon wavevector of transverse phonons of vertical
polarization for two values of the in-plane wavevector. For reference, the
form-factor of the plane-wave potential is provided. All form-factors are
normalized to its value for $q_z=0$. }
\end{figure}

\begin{figure}[tbp]
\caption{ (a) Schematics of the for measuring phonon emission form the
multiple quantum well structures. The phonon signal is registered by the
superconducting bolometer deposited on the back side of the substrate. (b)
Similar structure but with additional filter superlattice reflecting phonons
in narrow frequency bands. }
\end{figure}

\begin{figure}[tbp]
\caption{ The phonon emission spectrum of GaAs/AlAs MQW structure for
transverse (solid line) and longitudinal (dashed line) phonons. The values
of the angle between the emitted phonon wavevector and the MQW axis are
restricted by $\protect\theta_{max}=0.1$. The electron temperature is $100K$%
. }
\end{figure}

\begin{figure}
\caption{
Bolometer signals for on- and off-resonance optical pumping of sample A. The large peak is due to
transverse phonons. The time gate used for the measurements of the angle dependence of TA 
intensity is indicated.
}
\end{figure}

\begin{figure}
\caption{
Phonon image of sample B (difference between on- and off-resonance phonon intensities). 
The numbers indicate position of the laser spot with respect to bolometer in milimiters.
The horizontal dashed line show the axis of the linescans displayed in Fig. 7.
}
\end{figure}

\begin{figure}
\caption{
Linescans of the TA intensity as a function of emission angle for samples A and B.
}
\end{figure}


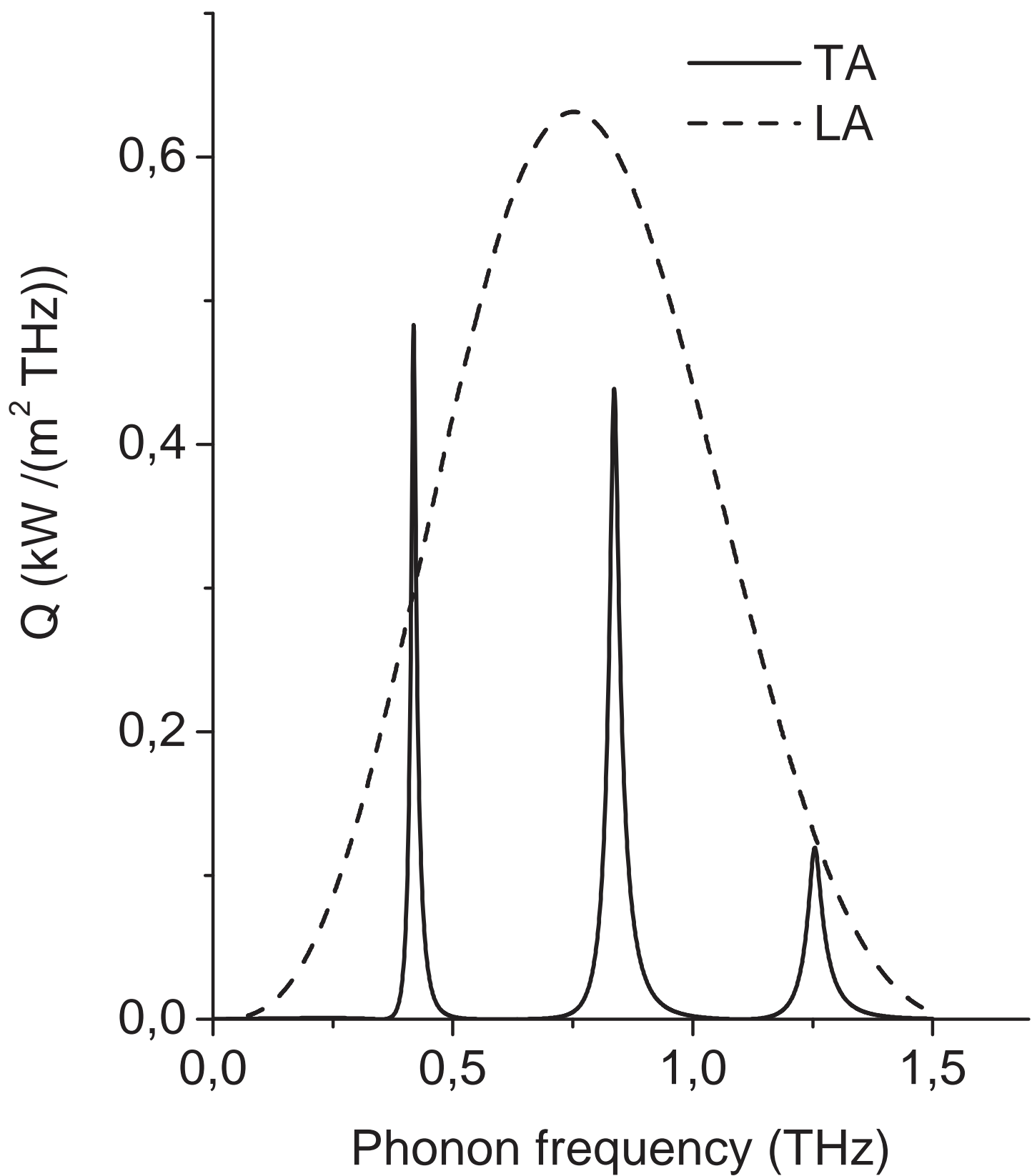
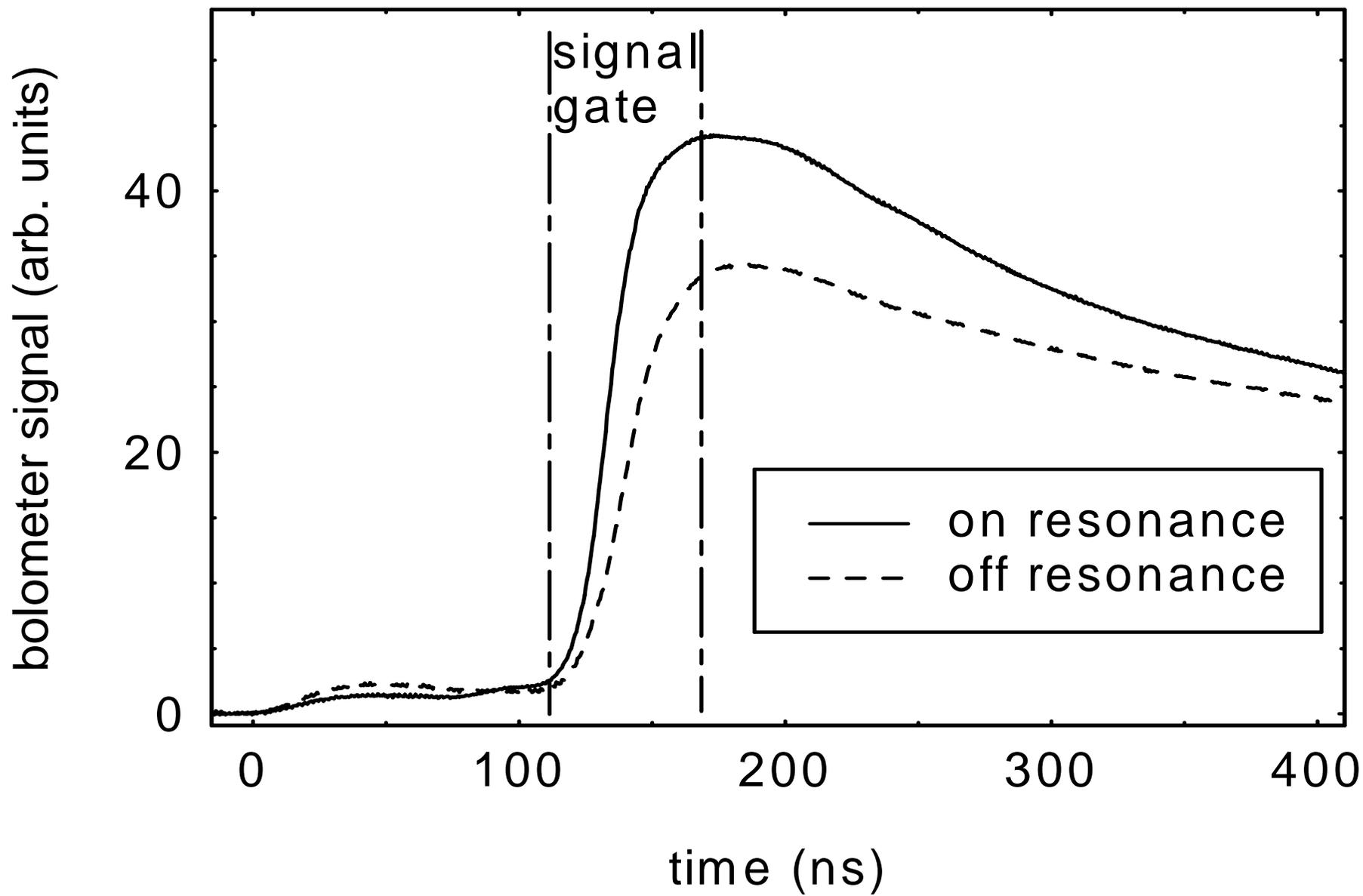
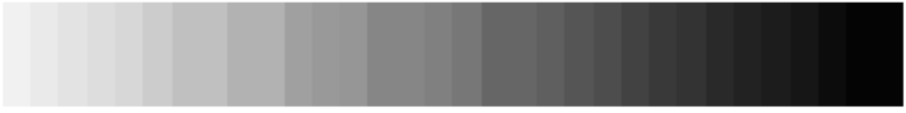
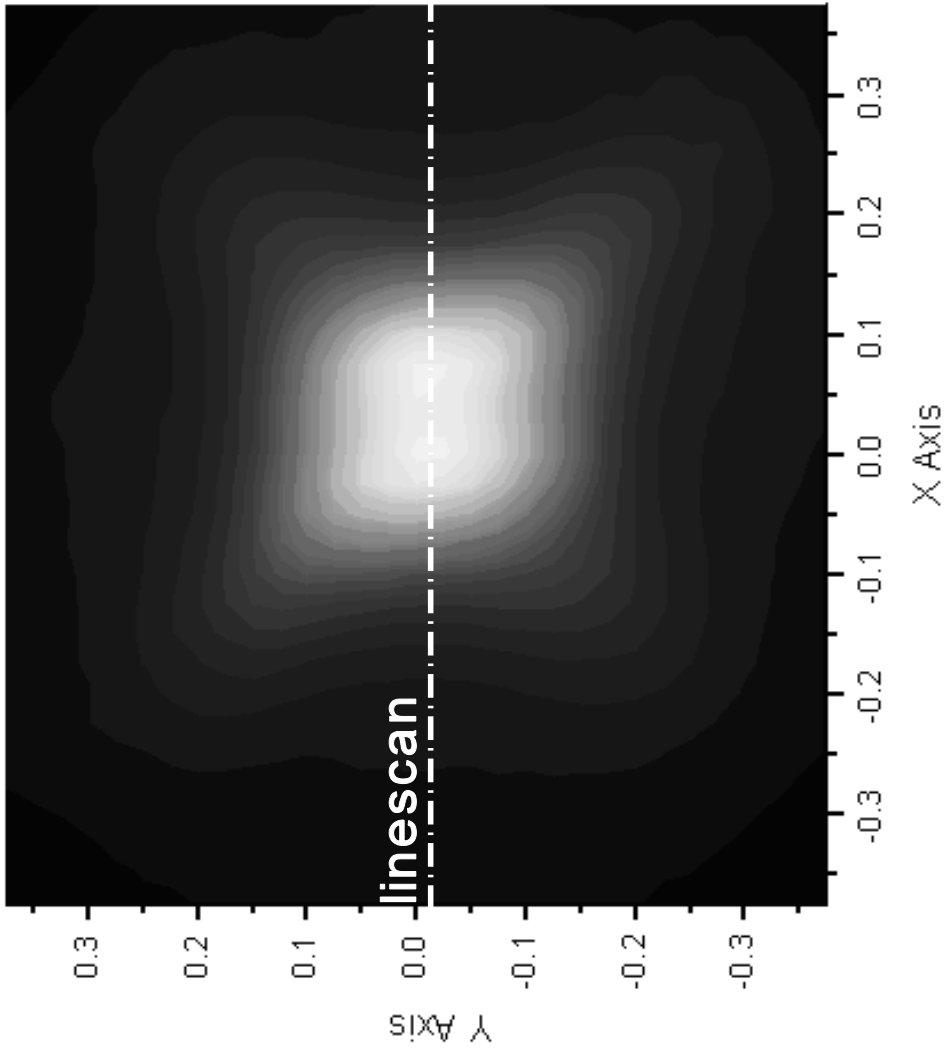
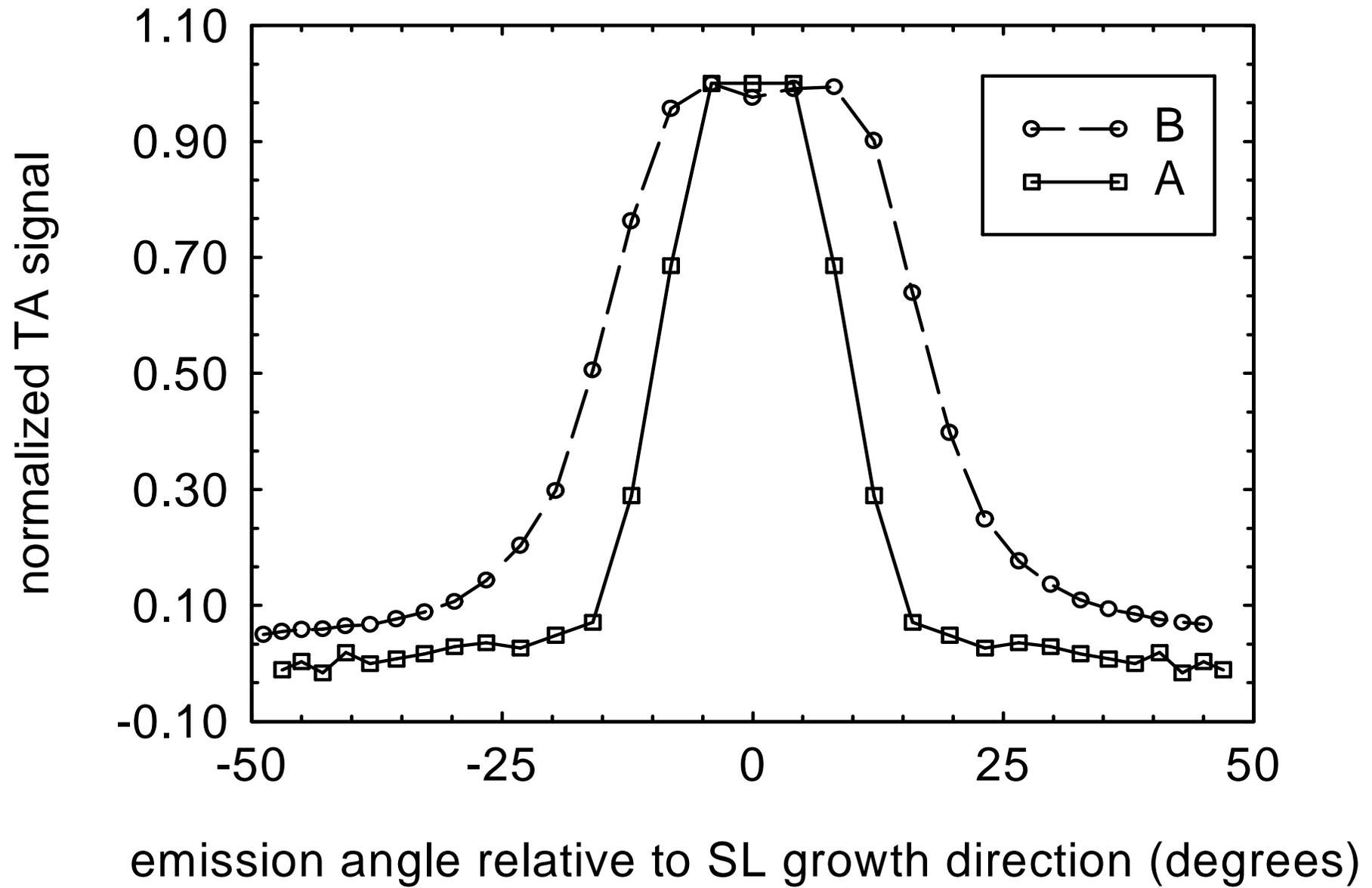
\end{document}